\documentclass[twocolumn,secnumarabic,amssymb, nobibnotes, aps, prc]{revtex4-2}

\long\gdef\COMMENT#1{}
\usepackage[russian,english]{babel}
\usepackage{xcolor}

\usepackage{amsmath}
\usepackage{graphicx}
\graphicspath{{images/}}
\usepackage{amsfonts}
\usepackage{amssymb}
\usepackage{color}
\usepackage{multirow}
\usepackage{babel}

\usepackage{lineno}

\begin{document}

\title{Finite Volume Effects on Transverse Momentum Spectra at LHC and RHIC Using a Blast-Wave Model with Planck Transformed Temperatures}

\author{A.S.~Parvan}
\email[e-mail: ]{parvan@theory.nipne.ro (corresponding author)}
\affiliation{BLTP, Joint Institute for Nuclear Research, Dubna, 141980, Russia}
\affiliation{DTP, Horia Hulubei National Institute for R$\&$D in Physics and Nuclear Engineering, M\u{a}gurele, Ilfov, 077125, Romania}

\author{A.A.~Aparin}
\affiliation{VBLHEP, Joint Institute for Nuclear Research, Dubna, 141980, Russia }
\affiliation {Moscow Institute of Physics and Technology, Dolgoprudny, 141700, Russia}

\author{E.V.~Nedorezov}
\affiliation{VBLHEP, Joint Institute for Nuclear Research, Dubna, 141980, Russia }
\affiliation {Moscow Institute of Physics and Technology, Dolgoprudny, 141700, Russia}

\begin{abstract}
We investigate finite volume effects on the transverse momentum spectra of charged pions produced in the most central heavy-ion collisions at RHIC and LHC energies. A cylindrically symmetric finite volume Boltzmann–Gibbs blast-wave model is employed that fully incorporates the finite longitudinal extent of the fire cylinder at kinetic freeze-out. The model applies Planck transformations to convert the local rest frame temperature and chemical potential of each fluid element into laboratory frame values, ensuring full Lorentz covariance. This approach is compared with the conventional infinite volume blast-wave model, in which the thermodynamic parameters remain defined in the local rest frame while the particle momenta are expressed in the laboratory frame. Both models are fitted to the experimental transverse momentum distributions of charged pions measured by the HADES, STAR, PHENIX, and ALICE collaborations over the center-of-mass energy range $  \sqrt{s_{NN}} = 2.4  $ GeV to $  5.44  $ TeV. The finite volume model with Planck transformed laboratory frame parameters yields temperature values fully consistent with relativistic thermodynamics (except for a small anomaly at $  \sqrt{s_{NN}} = 193  $ and $  200  $ GeV) and produces realistic fire cylinder volumes several times larger than the initial nuclear overlap volume. In contrast, the conventional infinite volume model yields unphysical results: infinite volume, infinite maximum half-length, and maximum longitudinal flow velocity equal to the speed of light at all energies. These findings demonstrate that a proper treatment of finite system size, together with the correct Lorentz (Planck form) transformation of the thermodynamic variables, is essential for the reliable extraction of freeze-out parameters in heavy-ion collisions.
\end{abstract}

\maketitle

\section{Introduction}
The blast-wave (BW) model, originally inspired by the classical hydrodynamic expansion of an exploding surface~\cite{Schnedermann1993,Siemens}, has emerged as one of the most successful phenomenological frameworks for describing particle production in high-energy heavy-ion collisions. It provides an excellent description of the transverse momentum ($  p_T  $) spectra and integrated yields of particles at kinetic freeze-out, as measured at the Relativistic Heavy Ion Collider (RHIC) and the Large Hadron Collider (LHC).
In its classical formulation~\cite{Siemens,Schnedermann1993,Retiere2004,Tang2009,Florkowski}, particles are assumed to be emitted from the surface of a radially boosted thermal source at the moment of kinetic freeze-out. The source is characterized by a common kinetic freeze-out temperature $  T_{\rm kin}  $ and a radially increasing flow velocity profile $  \beta(r)  $, which is typically assumed to rise linearly from the center to the edge of the fireball.

This simple yet powerful ansatz simultaneously reproduces several key experimental observations: the characteristic softening of pion spectra, the hardening of proton spectra, and the pronounced mass-dependent increase of the mean transverse momentum $  \langle p_T \rangle  $ as a function of collision centrality. These features are consistently described across a wide range of center-of-mass energies, from a few GeV up to $  \sqrt{s_{NN}} = 5.02  $ TeV~\cite{STAR2,ALICE3}.

Despite its phenomenological success, the majority of blast-wave fits performed by experimental collaborations (ALICE, STAR, PHENIX, and others) and in phenomenological studies still rely on an integration over an infinite emission volume. These fits typically do not properly normalize the particle yields to the finite size of the system~\cite{Adams2004,Adam2017,Adam2016,Tariq2024,Melo2020,Wang2020,Song2020,Sun2017,Ray2019,Liu2024,Saha2025,Ortiz2017}.

As a result, the extracted “temperature” and “radial flow velocity” parameters represent effective averaged quantities, while the system volume (or equivalently the overall normalization) is treated as a free external parameter. This normalization is usually fixed using particle ratios~\cite{Wang2020} or other external inputs, such as HBT radii~\cite{Tomasik2003}, rather than being determined intrinsically within the model itself.

This infinite volume approximation is physically unsatisfactory. At kinetic freeze-out, the geometric size of the fireball is expected to be only a few times larger than the initial nuclear overlap region, making a finite volume treatment both necessary and more appropriate.

Moreover, in the conventional blast-wave formalism~\cite{Siemens,Schnedermann1993,Retiere2004,Tang2009,Florkowski,Adams2004,Adam2017,Adam2016,Tariq2024,Melo2020,Wang2020,Song2020,Sun2017,Ray2019,Liu2024,Saha2025,Ortiz2017}, the thermodynamic parameters (temperature and chemical potential) are defined in the local rest frame of each fluid element on the emitting surface, while the particle momenta appearing in the distribution function are expressed in the laboratory (global) rest frame. In standard statistical mechanics, however, the particle momenta and the thermodynamic variables in the distribution function are defined in the same reference frame.

Furthermore, this definition in the conventional blast-wave model complicates direct comparisons with results from statistical models based on Tsallis non-extensive statistics~\cite{Cleymans13,Parvan2017,Azmi20}, which are formulated in the laboratory frame; with lattice QCD equation-of-state calculations~\cite{Bazavov2014,Borsanyi2014}, which are performed in the global rest frame of the fireball (i.e., the laboratory frame); and with thermodynamic parameters extracted from particle ratios using statistical thermal models, which are also formulated in the laboratory frame~\cite{BraunMunzinger09,Cleymans1999,Parvan2005}.

In this work, we employ an extended blast-wave framework recently introduced in Ref.~\cite{Parvan2024}. This approach explicitly incorporates the finite spatial extent of the emitting source at kinetic freeze-out and applies a proper Lorentz transformation -- referred to as the Planck transformation -- of the local thermodynamic variables (temperature and chemical potential) from the rest frame of each fluid element to the laboratory frame.

It should be noted that a finite volume, cylindrically symmetric blast-wave model, in which the temperature and chemical potential are defined in the rest frame of each fluid element while the particle momenta are specified in the laboratory frame, was originally introduced in Ref.~\cite{Schnedermann1993}. This approach has since been widely adopted in several studies~\cite{Rode2020,Nara2022,Choi2011}.

In this work, we compare the cylindrically symmetric finite volume Boltzmann–Gibbs blast-wave model -- which incorporates the Planck transformed laboratory frame temperature and chemical potential -- with the conventional infinite volume model (which employs rest frame temperature and chemical potential). Both models are fitted to the transverse-momentum spectra of charged pions produced in the most central heavy-ion collisions over a broad range of center-of-mass energies. In both cases, we study the energy dependence of the temperature, chemical potential, and transverse flow velocity.

Additionally, within the finite volume Boltzmann–Gibbs blast-wave model combined with the Planck transformation, we determine the energy dependence of the fire cylinder volume as well as other key quantities at kinetic freeze-out, including the maximum longitudinal flow velocity, the maximum half-length of the fire cylinder, and the maximum space-time rapidity. These quantities are inaccessible in the conventional infinite volume blast-wave model, where the fire cylinder volume, the maximum half length of the fire cylinder, and the maximum space-time rapidity remain infinite at all collision energies, while the maximum longitudinal flow velocity equals the speed of light.

The paper is organized as follows. In Sec.~2, we introduce the finite volume Boltzmann–Gibbs blast-wave model formulated on a freeze-out hypersurface with both spherical and cylindrical symmetry. Section~3 presents fits to the transverse momentum spectra of identified hadrons measured by the HADES, STAR, PHENIX, and ALICE collaborations in heavy-ion collisions. Finally, Sec.~4 summarizes the main conclusions and discusses the results.

\section{Finite Volume Blast-Wave Model}
In blast-wave models the fireball expands both longitudinally and transversely with a certain velocity distribution. Using this framework, the thermodynamic parameters of hot matter at the kinetic freeze-out stage have been extracted from the transverse momentum distributions of final particles over a wide range of collision energies and centralities. The BW model is based on the principle of local thermodynamic equilibrium, dividing the fireball into fluid elements, each treated as a thermodynamic system in equilibrium. Each fluid element moves with its own relativistic 3-velocity, $\mathbf{v}$, and its four-velocity is defined as $u^{\mu} = \gamma (1, \mathbf{v})$, where $\gamma = (1 - \mathbf{v}^2)^{-1/2}$ is the Lorentz factor. Each fluid element, modeled as an ideal gas of hadrons with four-momentum $p^{\mu} = (E_{\mathbf{p}}, \mathbf{p})$, is described within the grand canonical ensemble. For hadrons following Maxwell-Boltzmann statistics, the transverse momentum distribution in the Boltzmann-Gibbs blast-wave (BGBW) model is given by~\cite{Florkowski,Parvan2024}
\begin{equation}\label{1}
\frac{d^{2}N}{dp_{T}dy} = \frac{g}{(2\pi)^{3}} p_{T} \int d\Sigma_{\mu} p^{\mu}   \int\limits_{0}^{2\pi} d\varphi \ e^{-\frac{p_{\mu}u^{\mu}-\mu_{0}}{T_{0}}},
\end{equation}
where $d\Sigma_{\mu}$ represents the volume element of the freeze-out hypersurface, $p_{T}$, $y$, and $\varphi$ denote the particle's transverse momentum, rapidity, and azimuthal angle, respectively, and $T_{0}$ and $\mu_{0}$ are the temperature and chemical potential of the fluid element in its rest frame $K_{0}$, which moves at constant velocity $\mathbf{v}$ relative to the laboratory frame $K$. Here and in the following, we adopt natural units with $\hbar = c = k_B = 1$.

\subsection{Spherically Symmetric Blast-Wave Model}
We consider a spherically symmetric freeze-out hypersurface in the BGBW model~\cite{Siemens,Rischke}. Following the formalism in~\cite{Florkowski}, the Maxwell-Boltzmann local equilibrium transverse momentum distribution for a finite volume fireball is derived as:
\begin{align}\label{2}
\frac{d^{2}N}{dp_{T}dy} &= \frac{g e^{\frac{\mu_{0}}{T_{0}}}}{2\pi} \ p_{T}  m_{T} \cosh y \int\limits_{0}^{R_{s}} r^{2} dr \int\limits_{0}^{\pi} d\theta \sin\theta \nonumber \\
                        & \times  e^{-\frac{\gamma(r)}{T_{0}} m_{T} \left[ \cosh y - v(r) \sinh y \cos\theta \right] } \nonumber \\
                        & \times I_{0}\left(\frac{\gamma(r)}{T_{0}} p_{T} v(r) \sin\theta \right),
\end{align}
where $I_{0}(x)$ is the modified Bessel function of the first kind, $m_{T} = \sqrt{p_{T}^{2} + m^{2}}$ is the transverse mass, and $R_{s}$ is the radius of the fireball at freeze-out. The fluid element's velocity and Lorentz factor for a spherically symmetric hypersurface are:
\begin{equation}\label{3}
  v(r) = v_{\mathrm{f}} \ \frac{r}{R_{s}},  \qquad \gamma(r) = \frac{1}{\sqrt{1-v^{2}(r)}},
\end{equation}
where $v_{\mathrm{f}} = R_{s}/t_{\mathrm{f}}$ and $t_{\mathrm{f}}$ is the freeze-out time.

The fireball's volume at freeze-out is given by~\cite{Parvan2024}:
\begin{align}\label{4}
  V &\equiv \int d\Sigma^{0}  = \frac{4}{3} \pi R_{s}^{3}, \\ \label{5}
  d\Sigma^{0} &= \frac{dr}{d\zeta} r^{2}(\zeta) \sin\theta d\theta d\phi d\zeta,
\end{align}
where $\zeta \in [0, 1]$ is a parameterization variable, $r(\zeta) = \zeta R_{s}$, and $\theta$ and $\phi$ are the polar and azimuthal angles of the fluid element, respectively~\cite{Florkowski}.

The Planck transformations for temperature and chemical potential are~\cite{Haar,Parvan2024}:
\begin{equation}\label{6}
  T=\frac{T_{0}}{\gamma}, \qquad  \mu=\frac{\mu_{0}}{\gamma},
\end{equation}
where $T_{0}$ and $\mu_{0}$ are the temperature and chemical potential in the fluid element's rest frame, and $T$ and $\mu$ are those in the laboratory frame $K$.
Substituting Eq.~(\ref{6}) into Eq.~(\ref{2}), the transverse momentum distribution becomes:
\begin{align}\label{7}
\frac{d^{2}N}{dp_{T}dy} &= \frac{g e^{\frac{\mu}{T}}}{2\pi} \ p_{T}  m_{T} \cosh y \int\limits_{0}^{R_{s}} r^{2} dr \int\limits_{0}^{\pi} d\theta \sin\theta \nonumber \\
                        & \times  e^{-\frac{m_{T}}{T}  \left[ \cosh y - v(r) \sinh y \cos\theta \right] } \nonumber \\
                        & \times I_{0}\left(\frac{p_{T}}{T}  v(r) \sin\theta \right).
\end{align}

In the non-relativistic limit ($v(r) \to 0$), Eq.~(\ref{7}) simplifies to~\cite{Parvan2017}:
\begin{equation}\label{8}
\frac{d^{2}N}{dp_{T}dy} = \frac{gV}{(2\pi)^{2}} p_{T}  m_{T} \cosh y  \ e^{-\frac{m_{T} \cosh y-\mu}{T}},
\end{equation}
where $V$ is given by Eq.~(\ref{4}). This represents the Maxwell-Boltzmann global equilibrium transverse momentum distribution for hadrons in Boltzmann-Gibbs statistics.

\subsection{Cylindrically Symmetric Blast-Wave Model}
We consider a cylindrically symmetric hypersurface in the context of the blast-wave model, as introduced by Schnedermann et al.~\cite{Schnedermann1993}. This model describes particle production in high energy heavy-ion collisions at the freeze-out hypersurface, where particles cease interacting. Using the computational framework outlined by Florkowski~\cite{Florkowski}, we derive the Maxwell-Boltzmann local equilibrium transverse momentum distribution for the BGBW model for a finite volume freeze-out fire cylinder, as detailed in Ref.~\cite{Parvan2024}. The distribution is given by~\cite{Parvan2024}:
\begin{align}\label{t4}
\frac{d^{2}N}{dp_{T}dy} &= \frac{g e^{\frac{\mu_{0}}{T_{0}}}}{2\pi} \ p_{T}  m_{T} \frac{R_{c}}{\bar{v}_{T\mathrm{f}}} \int\limits_{0}^{R_{c}} r dr \int\limits_{-\eta_{\max}}^{\eta_{\max}} d\eta_{\|} \nonumber \\
                        & \times \cosh(\eta_{\|}-y) \ e^{-\frac{m_{T}}{T_{0}} \cosh\rho(r)\cosh(\eta_{\|}-y)} \nonumber \\
                        & \times I_{0}\left(\frac{p_{T}}{T_{0}} \sinh\rho(r) \right),
\end{align}
where \( g \) is the degeneracy factor, \( \mu_0 \) and \( T_0 \) are the chemical potential and temperature at freeze-out, \( p_T \) is the transverse momentum, \( m_T = \sqrt{p_T^2 + m^2} \) is the transverse mass (with \( m \) being the particle rest mass), \( y \) is the particle rapidity, and \( I_0 \) is the modified Bessel function of the first kind. The coordinates \( r \), \( \eta_{\parallel} \), and \( \phi \) represent the radial distance from the \( z \)-axis, spacetime rapidity, and azimuthal angle of the fluid element, respectively.

The hyperbolic functions \( \cosh\rho(r) \) and \( \sinh\rho(r) \) describe the Lorentz boost due to the transverse velocity of a fluid element at \( z = 0 \):
\begin{align}\label{t5}
  \cosh\rho(r) &=\frac{1}{\sqrt{1-\bar{v}_{T}^{2}(r)}}, \nonumber  \\
  \sinh\rho(r) &=\frac{\bar{v}_{T}(r)}{\sqrt{1-\bar{v}_{T}^{2}(r)}},
\end{align}
where the transverse velocity profile is:
\begin{equation}\label{t6}
  \bar{v}_{T}(r) = \bar{v}_{T\mathrm{f}} \ \frac{r}{R_{c}}.
\end{equation}
Here, \( \bar{v}_{T\mathrm{f}} = R_c / \tau_\mathrm{f} \) is the transverse velocity scale, \( R_c \) is the radius of the fire cylinder, and \( \tau_\mathrm{f} \) is the fixed proper time at freeze-out. The parameter \( \eta_{\mathrm{max}} \) denotes the maximal spacetime rapidity, bounding the longitudinal extent of the system.

The volume of the fire cylinder at freeze-out is defined as~\cite{Parvan2024}:
\begin{align}\label{t7}
  V &\equiv \int d\Sigma^{0}   = 2 z_{\max} \pi R_{c}^{2}, \\  \label{t8}
                  z_{\max} &= \tau_{\mathrm{f}} \sinh\eta_{\max} = \frac{R_{c}}{\bar{v}_{T\mathrm{f}}} \sinh\eta_{\max}, \\ \label{t8a}
  d\Sigma^{0} &= \frac{dr}{d\zeta} \tau(\zeta) r(\zeta) \cosh\eta_{\|} d\eta_{\|} d\phi d\zeta,
\end{align}
where \( r(\zeta) = \zeta R_c \), \( \tau(\zeta) = \tau_f =const \), and \( z_{\mathrm{max}} \) is the maximal half-length of the fire-cylinder along the \( z \)-axis, as described by Florkowski~\cite{Florkowski}.

In the limit \( \eta_{\mathrm{max}} \to \infty \), corresponding to an infinite volume (\( V \to \infty \)) of the fire cylinder (see Eqs.~\eqref{t7} and \eqref{t8}), Eq.~\eqref{t4} simplifies to the well-known transverse momentum distribution of the blast-wave model, as derived by Schnedermann et al.~\cite{Schnedermann1993}:
\begin{align}\label{t9}
\frac{d^{2}N}{dp_{T}dy} &= \frac{g e^{\frac{\mu_{0}}{T_{0}}}}{\pi} \ p_{T}  m_{T} \tau_{\mathrm{f}} \int\limits_{0}^{R_{c}} r dr K_{1}\left(\frac{m_{T}}{T_{0}} \cosh\rho(r)\right)  \nonumber \\
                        & \times  I_{0}\left(\frac{p_{T}}{T_{0}} \sinh\rho(r) \right).
\end{align}
where \( K_1 \) is the modified Bessel function of the second kind. This expression assumes an infinite longitudinal extent of the fire cylinder at freeze-out, which is atypical for heavy-ion collisions where finite volumes are more realistic. Moreover, in Eq.~(\ref{t9}), the thermodynamic parameters (temperature and chemical potential) are defined in the rest frame of the fluid element, while the particle momenta are specified in the laboratory frame. In statistical mechanics, the particle momenta and the thermodynamic variables in the distribution function are conventionally defined in the same reference frame.

The Lorentz factor, expressed in cylindrical parametrization, is given by~\cite{Florkowski,Parvan2024}
\begin{equation}\label{t11a}
  \gamma(r,\eta_{\|}) = \cosh\rho(r) \cosh\eta_{\|} = \frac{\cosh\eta_{\|}}{\sqrt{1-\bar{v}_{T}^{2}(r)}}.
\end{equation}

The distributions given in Eqs.~(\ref{t4}) and (\ref{t9}) define a constant temperature $  T_0  $ and chemical potential $  \mu_0  $ in the local rest frame of each fluid element. These values are identical for all fluid elements throughout the fireball. Consequently, the temperature $  T  $ and chemical potential $  \mu  $ in the laboratory frame for a given fluid element are obtained via the Planck transformations \eqref{6} and \eqref{t11a}:
\begin{align}\label{t11b}
  T(r,\eta_{\|},\phi) &= \frac{T_0 \sqrt{1-\bar{v}_{T}^{2}(r)}}{\cosh\eta_{\|}}, \nonumber \\
  \mu(r,\eta_{\|},\phi) &= \frac{\mu_0 \sqrt{1-\bar{v}_{T}^{2}(r)}}{\cosh\eta_{\|}},
\end{align}
where the radial coordinate satisfies $0 \leqslant r \leqslant R_c$, the azimuthal angle satisfies $0 \leqslant \phi < 2\pi$, and the space-time rapidity is constrained by $  |\eta_{\|} | \leqslant \eta_{\max}  $ in Eq.~(\ref{t4}) and by $  -\infty \leqslant \eta_{\|} \leqslant \infty  $ in Eq.~(\ref{t9}).

Applying the Planck transformations \eqref{6} and Eq.~\eqref{t11a} to Eq.~\eqref{t4}, we derive the Maxwell–Boltzmann local equilibrium transverse momentum distribution for the BGBW model with transformed temperature and chemical potential, as presented in Ref.~\cite{Parvan2024}:
\begin{align}\label{t12}
\frac{d^{2}N}{dp_{T}dy} &= \frac{g e^{\frac{\mu}{T}}}{2\pi} p_{T}  m_{T} \frac{R_{c}}{\bar{v}_{T\mathrm{f}}} \int\limits_{0}^{R_{c}} r dr \int\limits_{-\eta_{\max}}^{\eta_{\max}} d\eta_{\|} \nonumber \\
                        & \times \cosh(\eta_{\|}-y) \ e^{-\frac{m_{T}}{T} \frac{\cosh(\eta_{\|}-y)}{\cosh\eta_{\|}}} \nonumber \\
                        & \times I_{0}\left(\frac{p_{T}}{T} \frac{\bar{v}_{T}(r)}{\cosh\eta_{\|}} \right).
\end{align}
Here, \( T \) and \( \mu \) are the temperature and chemical potential in the laboratory reference frame \( K \), respectively, recalculated by the Planck transformations to account for relativistic effects. Integrating Eq.~\eqref{t12} with respect to the radial coordinate $r$, we obtain
\begin{align}\label{t12b}
\frac{d^{2}N}{dp_{T}dy} &= \frac{g e^{\frac{\mu}{T}}}{2\pi}  m_{T} \frac{T R_{c}^{3}}{\bar{v}_{T\mathrm{f}}^{2}} \int\limits_{-\eta_{\max}}^{\eta_{\max}} d\eta_{\|}  \cosh(\eta_{\|}-y) \nonumber \\
                        & \times \cosh(\eta_{\|})  e^{-\frac{m_{T}}{T} \frac{\cosh(\eta_{\|}-y)}{\cosh\eta_{\|}}}  I_{1}\left(\frac{p_{T}}{T} \frac{\bar{v}_{T\mathrm{f}}}{\cosh\eta_{\|}} \right),
\end{align}
where $I_1$ is the modified Bessel function of the first kind of order 1. The transverse momentum spectrum (Eq.~\eqref{t12b}) is Lorentz invariant and retains an explicit dependence on particle rapidity $y$. In contrast, the boost invariant spectrum of the infinite volume blast-wave model (Eq.~\eqref{t9}) is independent of rapidity $y$.

Equation~\eqref{t12} can be reformulated in terms of the transverse (\( v_T \)) and longitudinal (\( v_z \)) velocities of the fluid element:
\begin{align}\label{t13}
\frac{d^{2}N}{dp_{T}dy} &= \frac{g e^{\frac{\mu}{T}}}{2\pi} p_{T}  m_{T} \tau_{\mathrm{f}} \int\limits_{0}^{R_{c}} r dr \int\limits_{-\eta_{\max}}^{\eta_{\max}} d\eta_{\|} \nonumber \\
                        & \times \cosh(\eta_{\|}-y) \ e^{-\frac{m_{T}}{T} (\cosh y -v_{z}\sinh y) } \nonumber \\
                        & \times I_{0}\left(\frac{p_{T}v_{T}}{T}  \right),
\end{align}
where the velocity components are defined as (see Florkowski~\cite{Florkowski}):
\begin{equation}\label{t14}
  v_{T}(r,\eta_{\|})=\frac{\bar{v}_{T}(r)}{\cosh\eta_{\|}}, \qquad v_{z}(\eta_{\|}) =\tanh\eta_{\|}.
\end{equation}
In the limit \( \mathbf{v} \to 0 \) (i.e., \( v_T = 0 \), \( v_z = 0 \)), Eq.~\eqref{t13} reduces to the Maxwell-Boltzmann global equilibrium transverse momentum distribution of the Boltzmann-Gibbs statistics (Eq.~\eqref{8}), with the system volume \( V \) given by Eq.~\eqref{t7}.

\begin{figure}[!htb]
\minipage{0.49\textwidth}
\hspace{-0.3cm}
\includegraphics[width=\linewidth]{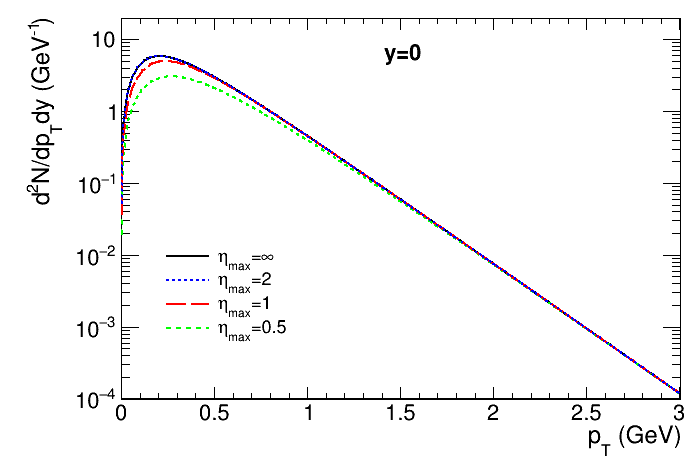}
\endminipage\hfill
\minipage{0.49\textwidth}
\vspace*{-0.66cm}
\includegraphics[width=\linewidth]{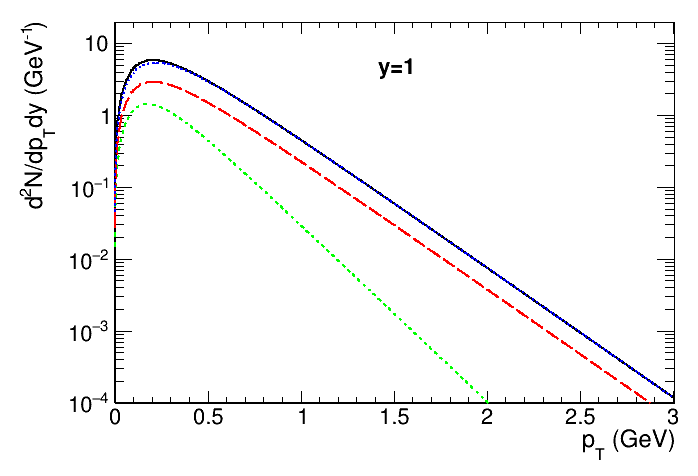}
\endminipage\hfill
\caption{(Color online) Transverse momentum distribution of positively charged pions $\pi^+$ in the blast-wave model with cylindrical symmetry for different values of maximal expansion rapidities $\eta_{\mathrm{max}} = 0.5, 1, 2$, and $\infty$ at particle rapidities $y = 0$ (top) and $y = 1$ (bottom). Calculated using Eqs.~\eqref{t4} and \eqref{t9} for a fire cylinder at freeze-out with temperature $T_0 = 120$ MeV and chemical potential $\mu_0 = 10$ MeV at the rest frame, radius $R_c = 3$ fm, and transverse flow velocity $\bar{v}_{T\mathrm{f}} = 0.6$.}
\label{fig1}
\end{figure}

\subsection{Cylindrical vs. Spherical Symmetry in Blast-Wave Models}
We compare the transverse momentum distributions of positively charged pions $\pi^+$ in the BGBW model under finite volume and infinite volume assumptions with cylindrical symmetry. Figure~\ref{fig1} illustrates these distributions for a fire cylinder at freeze-out, comparing the infinite volume case (Eq.~\eqref{t9}) with the finite volume case (Eq.~\eqref{t4}) at rapidities $y = 0$ (top) and $y = 1$ (bottom). The calculations use temperature $T_0 = 120$ MeV and chemical potential $\mu_0 = 10$ MeV at the rest frame, fire-cylinder radius $R_c = 3$ fm, and transverse flow velocity $\bar{v}_{T\mathrm{f}} = 0.6$. The dashed, long-dashed, dotted, and solid lines correspond to $\eta_{\mathrm{max}} = 0.5, 1, 2$, and $\infty$, respectively.

\begin{figure}[!htb]
\minipage{0.49\textwidth}
\includegraphics[width=\linewidth]{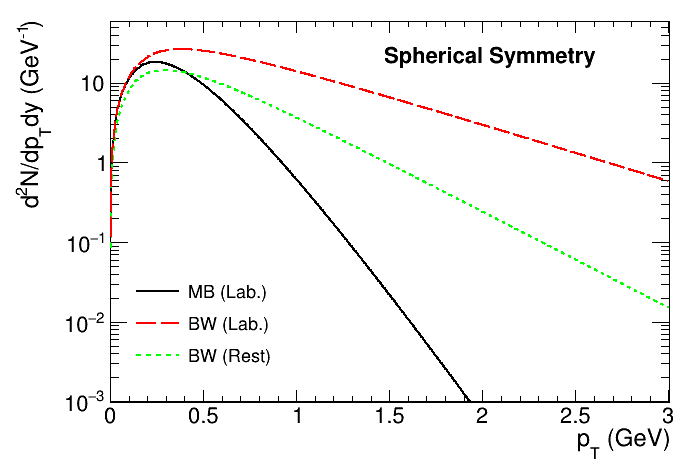}
\endminipage\hfill
\minipage{0.49\textwidth}
\vspace*{-0.66cm}
\hspace{-0.2cm}
\includegraphics[width=\linewidth]{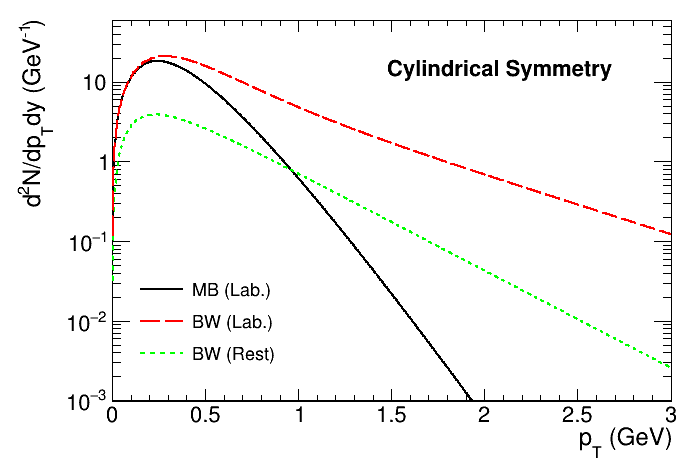}
\endminipage\hfill
\caption{(Color online) Comparison of Maxwell-Boltzmann transverse momentum distributions for positively charged pions $\pi^+$ at rapidity $y = 0$ in the blast-wave model under spherical (top) and cylindrical (bottom) symmetries. Solid lines depict the global equilibrium Maxwell-Boltzmann distribution, while dashed lines represent BW distributions calculated via Planck transformations using laboratory frame parameters at $T = 120$ MeV and $\mu = 10$ MeV at the local fluid cell frame. Dotted lines show BW distributions in the rest frame of a fluid element, evaluated at $T_0 = 120$ MeV and $\mu_0 = 10$ MeV. For the spherical fireball, $R_s = 5.68$ fm and $v_{\mathrm{f}} = 0.8$. For the cylindrical fire-cylinder, $R_c = 3$ fm, $\eta_{\mathrm{max}} = 2$, and $\bar{v}_{T\mathrm{f}} = 0.8$.}
\label{fig2}
\end{figure}

The finite volume BW model’s transverse momentum distribution deviates from the infinite volume case, with the discrepancy increasing at higher rapidity $y$ (see Figure~\ref{fig1}, top and bottom panels). In heavy-ion collisions, the fire cylinder at freeze-out has a finite volume, rendering the infinite volume BW model (Eq.~\eqref{t9}) potentially unreliable especially at higher particle rapidities, where the boost-invariance breaks down. The infinite volume model is boost-invariant, producing a transverse momentum distribution independent of rapidity $y$ due to $\eta_{\mathrm{max}} = \infty$, which assumes an infinitely long fire-cylinder which is not the case in realistic scenarios. In contrast, the finite volume BW model (Eq.~\eqref{t4}) reflects the finite length of fire-cylinders in heavy-ion collisions, resulting in a non-boost-invariant distribution that varies with rapidity $y$.

Figure~\ref{fig2} compares the transverse momentum distributions of the BW model with spherical (top) and cylindrical (bottom) geometry of the fireball expansion for $\pi^+$ pions at mid-rapidity $y = 0$. It also compares the Maxwell-Boltzmann global equilibrium distribution with the Maxwell-Boltzmann local equilibrium distributions derived from the BW model. These distributions use either local parameters defined at the rest frame of a fluid element (Eqs.~\eqref{2} and \eqref{t4}) or global parameters defined at the laboratory frame via Planck transformations (Eqs.~\eqref{7} and \eqref{t12b}). Calculations employ $T = 120$ MeV and $\mu = 10$ MeV for global parameters (Eqs.~\eqref{7}, \eqref{8}, \eqref{t12b}) and $T_0 = 120$ MeV and $\mu_0 = 10$ MeV for local parameters (Eqs.~\eqref{2}, \eqref{t4}). For the spherical fireball, we use $R_s = 5.68$ fm and $v_{\mathrm{f}} = 0.8$; for the fire cylinder, $R_c = 3$ fm, $\eta_{\mathrm{max}} = 2$, and $\bar{v}_{T\mathrm{f}} = 0.8$, with equal volumes for both geometries.

The Planck transformations align the local equilibrium transverse momentum distributions for both spherical and cylindrical symmetry cases with the Maxwell-Boltzmann global equilibrium distribution in the laboratory frame. However, the local equilibrium distributions calculated using local parameters in the fluid element’s rest frame do not match the global equilibrium distribution at the same temperature and chemical potential. Figure~\ref{fig2} demonstrates that the fireball’s spatial symmetry significantly influences the transverse momentum distribution.

\begin{table*}
\caption{Fit parameters of the finite volume Boltzmann-Gibbs blast-wave model with Planck transformations for pions produced in the most central heavy ion collisions at various energies.}
\label{t1}
\begin{tabular}{cccccccccc}
 \hline
 \hline
 Collaboration  & Type   & Reaction & $\sqrt{s_{NN}}$, GeV & $\quad$ $T$, MeV $\qquad$  & $\quad$ $\mu$, MeV $\qquad$ & $\qquad$ $\bar{v}_{T\mathrm{f}}$ $\qquad$ &  $\qquad$ $\eta_{\mathrm{max}}$ $\qquad$ &  $\chi^{2}/ndf$  \\
 \hline
 HADES       & $\pi^{-}$ & Au+Au  & 2.4      & 30.98$\pm$2.42   & 159.59$\pm$1.83   & 0.698$\pm$0.014      & 1.759$\pm$0.077       & 2.54/18    \\
 STAR        & $\pi^{+}$ & Au+Au  & 7.7      & 55.51$\pm$31.80  & 250.01$\pm$59.81  & 0.777$\pm$0.161      & 0.779$\pm$0.131       & 5.49/22     \\
 STAR        & $\pi^{+}$ & Au+Au  & 11.5     & 40.39$\pm$24.63  & 227.10$\pm$130.71 & 0.845$\pm$0.076     & 1.000$\pm$0.230       & 1.28/22    \\
 STAR        & $\pi^{+}$ & Au+Au  & 19.6     & 56.91$\pm$48.25  & 267.62$\pm$65.30  & 0.796$\pm$0.155      & 1.063$\pm$0.211       & 0.21/22    \\
 STAR        & $\pi^{+}$ & Au+Au  & 27       & 71.52$\pm$21.40  & 282.38$\pm$9.81   & 0.756$\pm$0.109      & 1.184$\pm$0.340       & 0.94/22    \\
 STAR        & $\pi^{+}$ & Au+Au  & 39       & 69.33$\pm$21.46  & 282.16$\pm$18.07  & 0.773$\pm$0.103      & 1.178$\pm$0.297       & 0.55/22    \\
 STAR        & $\pi^{+}$ & Au+Au  & 62.4     & 49.56$\pm$27.21  & 272.59$\pm$65.55  & 0.829$\pm$0.090      & 0.961$\pm$0.129       & 0.40/6    \\
 STAR        & $\pi^{+}$ &  U+U   & 193      & 75.06$\pm$20.37  & 321.91$\pm$13.57  & 0.779$\pm$0.055      & 1.300$\pm$0.253       & 0.47/24    \\
 PHENIX      & $\pi^{+}$ & Au+Au  & 200      & 78.21$\pm$3.39   & 313.96$\pm$1.15   & 0.768$\pm$0.009      & 1.350$\pm$0.043       & 33.39/24    \\
 ALICE       & $\pi^{+}$ & Pb+Pb  & 2760     & 52.86$\pm$10.00  & 216.91$\pm$10.28  & 0.869$\pm$0.023      & 1.595$\pm$0.065       & 6.51/37    \\
 ALICE & $\pi^{-}+\pi^{+}$ & Pb+Pb & 2760    & 50.60$\pm$12.19  & 173.43$\pm$2.87   & 0.862$\pm$0.032      & 1.877$\pm$0.081       & 7.81/38    \\
 ALICE & $\pi^{-}+\pi^{+}$ & Pb+Pb & 5020    & 72.86$\pm$5.10   & 238.06$\pm$1.56   & 0.836$\pm$0.010      & 1.545$\pm$0.055       & 17.90/33    \\
 ALICE & $\pi^{-}+\pi^{+}$ & Xe+Xe & 5440    & 47.27$\pm$7.95   & 211.83$\pm$11.59  & 0.890$\pm$0.017      & 1.561$\pm$0.054       & 10.19/32    \\
\hline
\hline
\end{tabular}
\end{table*}

\begin{table*}
\caption{Fit parameters of the infinite volume Boltzmann-Gibbs blast-wave model with rest frame parameters for pions produced in the most central heavy ion collisions at various energies.}
\label{t2}
\begin{tabular}{cccccccccc}
 \hline
 \hline
 Collaboration  & Type   & Reaction & $\sqrt{s_{NN}}$, GeV & $\quad$ $T_{0}$, MeV $\qquad$  & $\quad$ $\mu_{0}$, MeV $\qquad$ & $\qquad$ $\bar{v}_{T\mathrm{f}}$ $\qquad$ &    $\chi^{2}/ndf$  \\
 \hline
 STAR        & $\pi^{+}$ & Au+Au  & 7.7      & 113.55$\pm$16.14 & 322.65$\pm$26.23  & 0.653$\pm$0.087            & 4.82/23     \\
 STAR        & $\pi^{+}$ & Au+Au  & 11.5     & 88.33$\pm$14.96  & 368.15$\pm$2.13   & 0.802$\pm$0.076            & 1.34/23    \\
 STAR        & $\pi^{+}$ & Au+Au  & 19.6     & 76.45$\pm$14.46  & 383.89$\pm$14.77  & 0.863$\pm$0.058            & 0.24/23    \\
 STAR        & $\pi^{+}$ & Au+Au  & 27       & 75.07$\pm$13.81  & 388.53$\pm$16.30  & 0.875$\pm$0.051            & 1.22/23    \\
 STAR        & $\pi^{+}$ & Au+Au  & 39       & 74.50$\pm$13.34  & 390.84$\pm$16.82  & 0.887$\pm$0.045            & 0.64/23    \\
 STAR        & $\pi^{+}$ & Au+Au  & 62.4     & 100.19$\pm$18.96 & 430.70$\pm$5.55   & 0.784$\pm$0.101            & 0.32/7    \\
 STAR        & $\pi^{+}$ &  U+U   & 193      & 73.48$\pm$11.44  & 432.70$\pm$21.24  & 0.910$\pm$0.030            & 0.87/25    \\
 PHENIX      & $\pi^{+}$ & Au+Au  & 200      & 71.10$\pm$1.93   & 419.78$\pm$3.49   & 0.914$\pm$0.005            & 54.6/25    \\
 ALICE       & $\pi^{+}$ & Pb+Pb  & 2760     & 75.07$\pm$5.24   & 367.01$\pm$4.53   & 0.947$\pm$0.008            & 7.51/38    \\
 ALICE & $\pi^{-}+\pi^{+}$ & Pb+Pb & 2760    & 74.75$\pm$4.88   & 331.55$\pm$2.06   & 0.948$\pm$0.007            & 9.51/39    \\
 ALICE & $\pi^{-}+\pi^{+}$ & Pb+Pb & 5020    & 92.34$\pm$2.55   & 393.34$\pm$1.49   & 0.930$\pm$0.004            & 15.91/34    \\
 ALICE & $\pi^{-}+\pi^{+}$ & Xe+Xe & 5440    & 79.32$\pm$3.42   & 375.85$\pm$3.12   & 0.950$\pm$0.005            & 14.83/33    \\
\hline
\hline
\end{tabular}
\end{table*}

\section{Analysis and results}
The experimental transverse momentum spectra of charged pions produced in $A-A$ collisions over a wide range of collision energy $\sqrt{s_{NN}} = 2.4$ GeV to $5.44$ TeV were fitted using two variants of the cylindrically symmetric BGBW model: (i) the BGBW model for a finite volume fire cylinder $(\eta_{\max} < \infty)$, incorporating Planck transformations of temperature and chemical potential; (ii) the standard BGBW model for an infinite volume fire-cylinder $(\eta_{\max} = \infty)$, employing rest frame temperature and chemical potential. Fits to the final particle spectra were performed using the ROOT analysis framework with the Minuit/Migrad minimization algorithm, taking into account both statistical and systematic uncertainties. The transverse momentum distributions [Eqs.~(\ref{t12b}) and (\ref{t9})] were integrated over particle rapidity $y$ within the experimental acceptance $y_0 \leqslant y \leqslant y_1$. The rapidity-integrated spectrum for the finite volume model [from Eq.~(\ref{t12b})] is:
\begin{align}\label{t12a}
\left.\frac{d^{2}N}{dp_{T}dy}\right|_{y_{0}}^{y_{1}} &= \frac{g e^{\frac{\mu}{T}}}{2\pi}  m_{T} \frac{R_{c}^{3}T}{\bar{v}_{T\mathrm{f}}^{2}} \int\limits_{y_{0}}^{y_{1}} dy \int\limits_{-\eta_{\max}}^{\eta_{\max}} d\eta_{\|} \cosh(\eta_{\|}) \nonumber \\
                        & \times  \cosh(\eta_{\|}-y) \  e^{-\frac{m_{T}}{T} \frac{\cosh(\eta_{\|}-y)}{\cosh\eta_{\|}}} \nonumber \\
                        & \times I_{1}\left(\frac{p_{T}}{T} \frac{\bar{v}_{T\mathrm{f}}}{\cosh\eta_{\|}} \right),
\end{align}
where $\bar{v}_{T\mathrm{f}}$ is the transverse flow velocity scale defined in Eq.~(\ref{t6}). The corresponding integrated spectrum for the infinite volume model [from Eq.~(\ref{t9})] is:
\begin{align}\label{t9a}
\left.\frac{d^{2}N}{dp_{T}dy}\right|_{y_{0}}^{y_{1}} &= \frac{g e^{\frac{\mu_{0}}{T_{0}}}}{\pi} \ p_{T}  m_{T}  \frac{R_{c}}{\bar{v}_{T\mathrm{f}}} \int\limits_{y_{0}}^{y_{1}} dy  \int\limits_{0}^{R_{c}} r dr   \nonumber \\
                        & \times   K_{1}(\frac{m_{T}}{T_{0}} \cosh\rho(r))  I_{0}(\frac{p_{T}}{T_{0}} \sinh\rho(r) ),
\end{align}
with the boost functions $\cosh\rho(r)$ and $\sinh\rho(r)$ given by Eq.~(\ref{t5}).

Equation~(\ref{t12a}) involves five fit parameters, while Eq.~(\ref{t9a}) has four. However, the chemical potential ($\mu$ or $\mu_0$) and cylinder radius $R_c$ are strongly correlated in both cases. To reduce degeneracy and ensure physical consistency, $R_c$ is fixed to the nuclear radius $R_A = r_0 A^{1/3}$ with $r_0 = 1.2$ fm throughout the analysis.

\begin{figure}[htbp]
\includegraphics[width=0.49\textwidth]{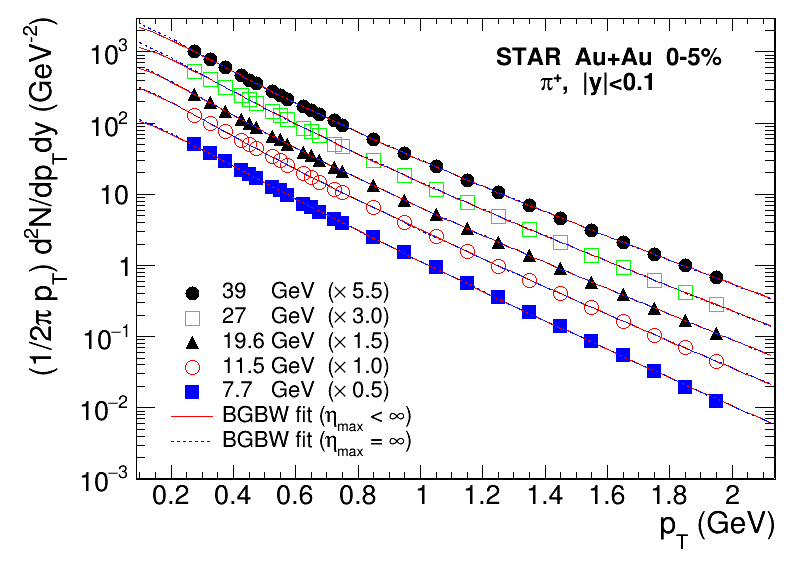}
\caption{(Color online) Transverse momentum spectra of positively charged pions $\pi^+$ produced in central Au+Au collisions at $\sqrt{s_{NN}} = 7.7$, $11.5$, $19.6$, $27$, and $39$ GeV, measured by the STAR Collaboration~\cite{STAR1} within $|y| < 0.1$. Solid curves show fits using the finite volume BGBW model ($\eta_{\max} < \infty$) with Planck transformation of temperature and chemical potential. Dashed curves represent fits from the standard infinite volume BGBW model ($\eta_{\max} = \infty$) using the rest frame parameters. Symbols denote experimental data points. Numbers indicate the scaling factors applied.}
\label{fig3}
\end{figure}

\begin{figure}[htbp]
\includegraphics[width=0.49\textwidth]{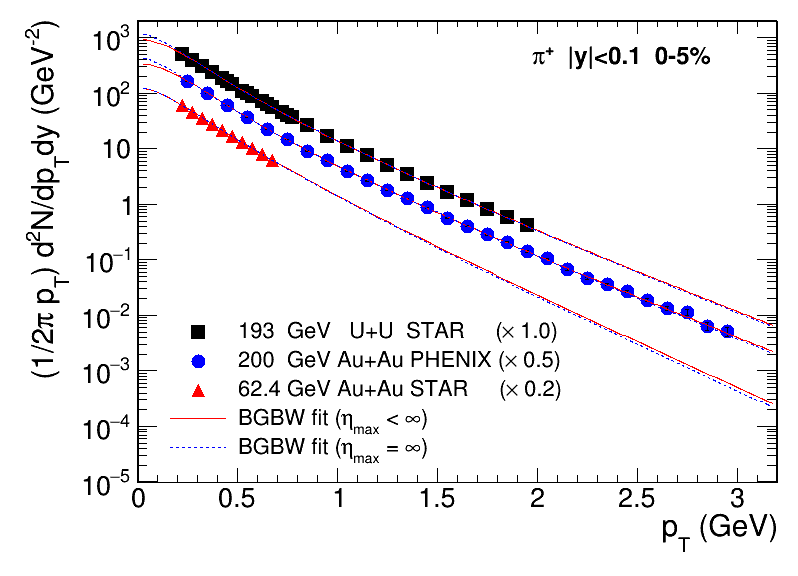}
\caption{(Color online) Transverse momentum spectra of positively charged pions $\pi^+$ measured by the STAR Collaboration in central Au+Au~\cite{STAR2} and U+U~\cite{STAR3} collisions at center-of-mass energies $\sqrt{s_{NN}} = 62.4$ and $193$ GeV, respectively, within the rapidity range $|y| < 0.1$, and by the PHENIX Collaboration~\cite{PHENIX1} in central Au+Au collisions at $\sqrt{s_{NN}} = 200$ GeV at mid-rapidity. The curve notations follow the same convention as in Figure~\ref{fig3}. Symbols indicate experimental data points. Numbers denote applied scaling factors.}
\label{fig4}
\end{figure}

\begin{table*}
\caption{ Parameters calculated using the finite volume Boltzmann-Gibbs blast-wave model with Planck transformation for pions produced in the most central heavy-ion collisions at various energies.}
\label{t3}
\begin{tabular}{ccccccc}
 \hline
 \hline
 Collaboration  & Type   & Reaction & $\sqrt{s_{NN}}$, GeV & $\qquad$ $v_{z,\mathrm{max}}$ $\qquad$  & $\qquad$ $z_{\mathrm{max}}$, fm $\qquad$ & $\qquad$ $V$, fm$^{3}$ $\qquad\quad$   \\
 \hline
 HADES       & $\pi^{-}$ & Au+Au  & 2.4      & 0.942$\pm$0.009  & 28.18$\pm$2.36   & 8627.81$\pm$721.38     \\
 STAR        & $\pi^{+}$ & Au+Au  & 7.7      & 0.652$\pm$0.075  & 7.74$\pm$2.23    & 2367.84$\pm$684.06       \\
 STAR        & $\pi^{+}$ & Au+Au  & 11.5     & 0.762$\pm$0.096  & 9.71$\pm$3.06    & 2972.67$\pm$935.38      \\
 STAR        & $\pi^{+}$ & Au+Au  & 19.6     & 0.787$\pm$0.080  & 11.19$\pm$3.71   & 3424.03$\pm$1136.37      \\
 STAR        & $\pi^{+}$ & Au+Au  & 27       & 0.829$\pm$0.106  & 13.67$\pm$5.94   & 4185.29$\pm$1818.99      \\
 STAR        & $\pi^{+}$ & Au+Au  & 39       & 0.827$\pm$0.094  & 13.27$\pm$5.09   & 4062.35$\pm$1556.86     \\
 STAR        & $\pi^{+}$ & Au+Au  & 62.4     & 0.745$\pm$0.057  & 9.40$\pm$1.92    & 2876.19$\pm$588.06      \\
 STAR        & $\pi^{+}$ &  U+U   & 193      & 0.862$\pm$0.065  & 16.22$\pm$4.90   & 5640.80$\pm$1704.13     \\
 PHENIX      & $\pi^{+}$ & Au+Au  & 200      & 0.874$\pm$0.010  & 16.33$\pm$0.83   & 4998.66$\pm$252.96      \\
 ALICE       & $\pi^{+}$ & Pb+Pb  & 2760     & 0.921$\pm$0.010  & 19.32$\pm$1.46   & 6137.25$\pm$462.79      \\
 ALICE & $\pi^{-}+\pi^{+}$ & Pb+Pb & 2760    & 0.954$\pm$0.007  & 26.31$\pm$2.44   & 8356.07$\pm$774.54      \\
 ALICE & $\pi^{-}+\pi^{+}$ & Pb+Pb & 5020    & 0.913$\pm$0.009  & 19.03$\pm$1.17   & 6045.65$\pm$372.99     \\
 ALICE & $\pi^{-}+\pi^{+}$ & Xe+Xe & 5440    & 0.916$\pm$0.009  & 15.65$\pm$0.96   & 3670.10$\pm$225.84     \\
\hline
\hline
\end{tabular}
\end{table*}

\begin{figure}[htbp]
\includegraphics[width=0.49\textwidth]{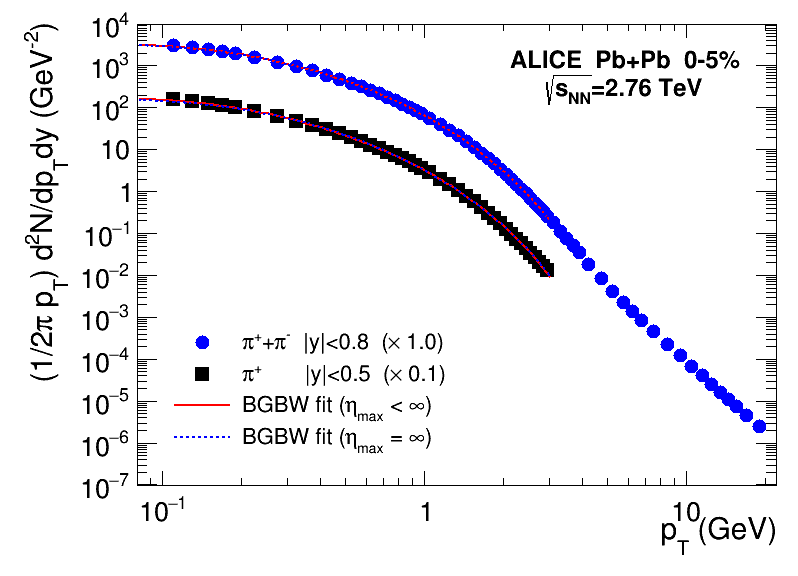}
\caption{(Color online) Transverse momentum spectra of $\pi^+$ and combined $\pi^+ + \pi^-$ measured by the ALICE Collaboration~\cite{ALICE1,ALICE2} in central Pb+Pb collisions at a center-of-mass energy of $\sqrt{s_{NN}} = 2.76$ TeV, within rapidity ranges $|y| < 0.5$ and $|y| < 0.8$, respectively. The curve notations are the same as in Figure~\ref{fig3}. Symbols denote experimental data points, and numbers indicate applied scaling factors.}
\label{fig5}
\end{figure}

\begin{figure}[htbp]
\includegraphics[width=0.49\textwidth]{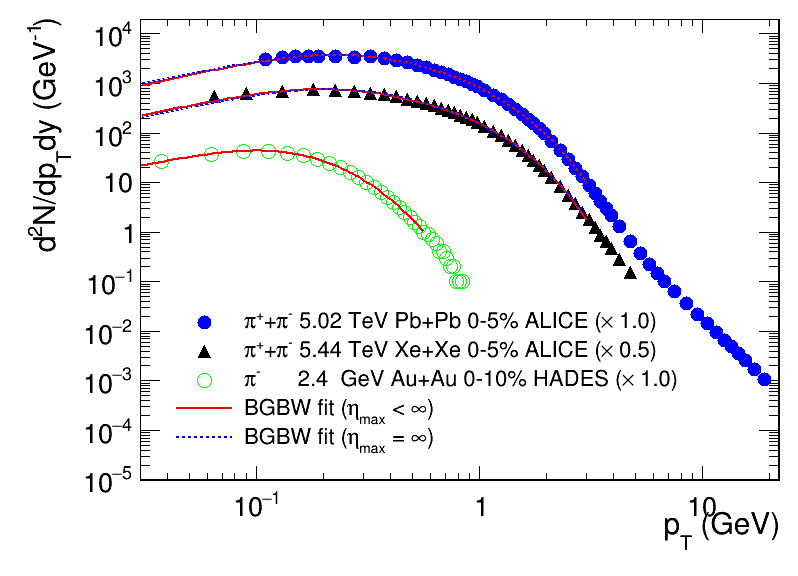}
\caption{(Color online) Transverse momentum spectra of charged pions $\pi^+ + \pi^-$ measured by the ALICE Collaboration in central Pb+Pb~\cite{ALICE3} and Xe+Xe~\cite{ALICE4} collisions at mid-rapidity $|y| < 0.5$ with center-of-mass energies $\sqrt{s_{NN}} = 5.02$ TeV and $5.44$ TeV, respectively, and of negatively charged $\pi^-$ measured by the HADES Collaboration~\cite{HADES1} in central Au+Au collisions at $\sqrt{s_{NN}} = 2.4$ GeV at mid-rapidity. The curve notations are the same as in Figure~\ref{fig3}. Symbols indicate experimental data points, with numerical labels denoting applied scaling factors.}
\label{fig6}
\end{figure}

\begin{figure*}[htbp]
\minipage{0.49\textwidth}
\includegraphics[width=\textwidth]{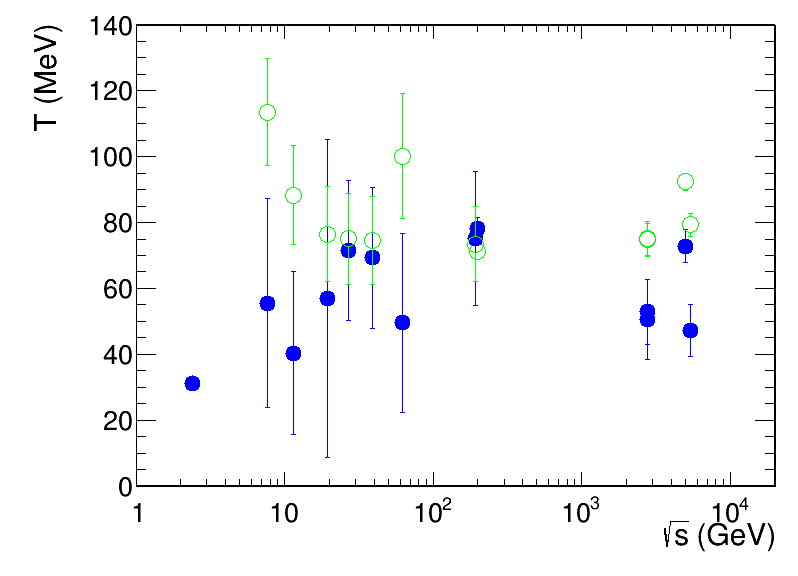}
\endminipage\hfill
\hspace{0.0\textwidth}
\minipage{0.49\textwidth}
\includegraphics[width=\textwidth]{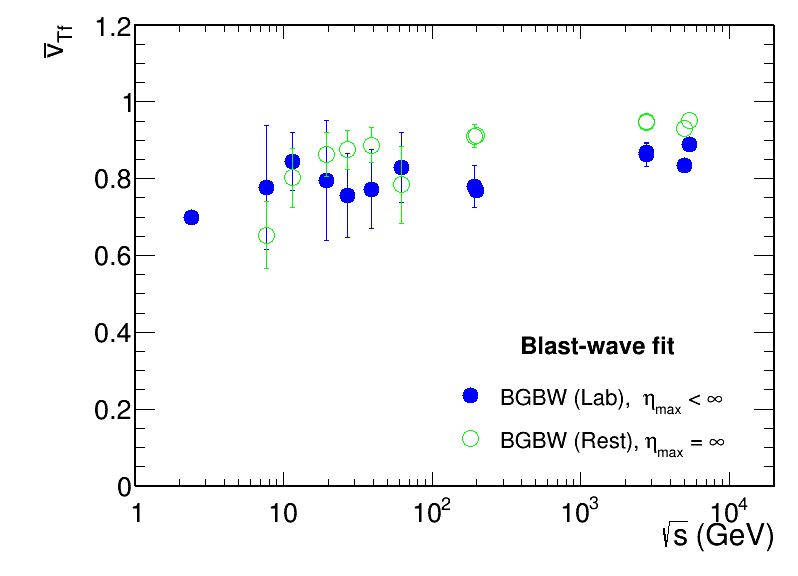}
\endminipage\hfill
\minipage{0.49\textwidth}
\includegraphics[width=\textwidth]{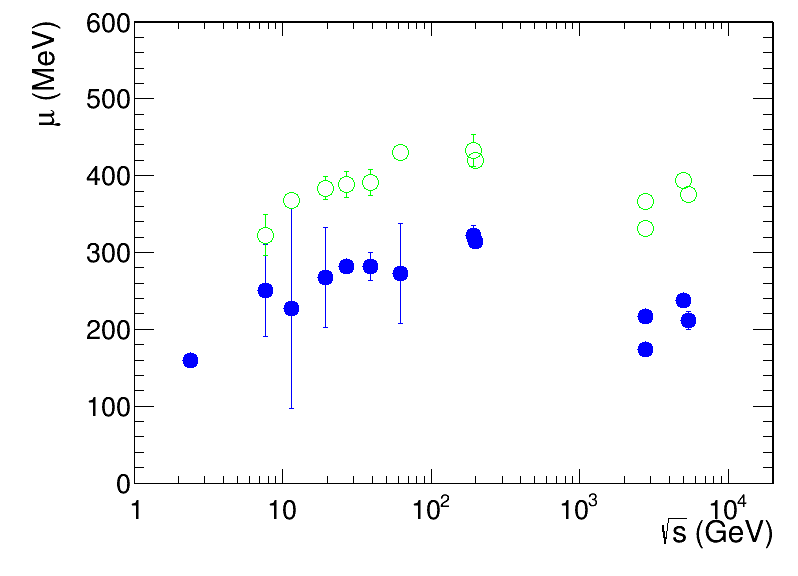}
\endminipage\hfill
\hspace{0.0\textwidth}
\minipage{0.49\textwidth}
\includegraphics[width=\textwidth]{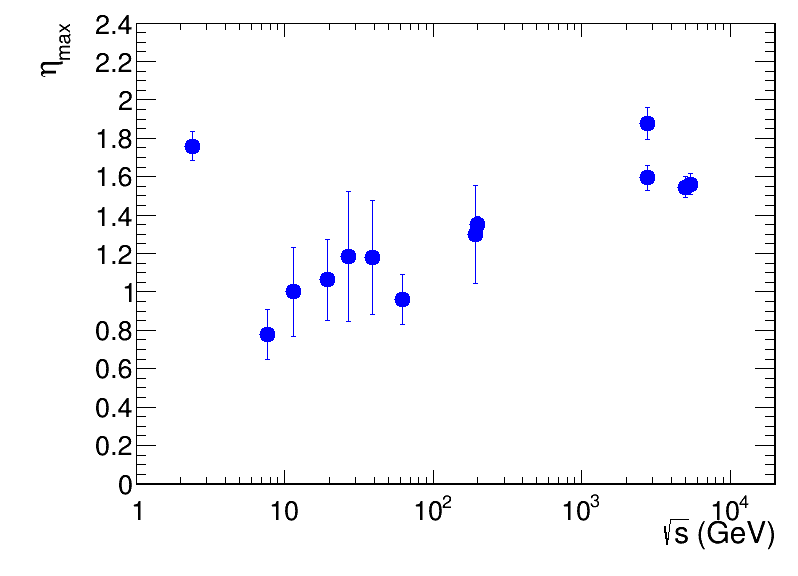}
\endminipage\hfill
\caption{(Color online) Energy dependence of the freeze-out parameters — temperature $T$, chemical potential $\mu$, transverse flow velocity $\bar{v}_{T\mathrm{f}}$, and maximal spacetime rapidity $\eta_{\mathrm{max}}$ — obtained from Maxwell-Boltzmann fits to transverse momentum particle spectra. Solid circles represent the finite volume Boltzmann-Gibbs blast-wave model with Planck transformed temperature and chemical potential, while open circles correspond to the infinite volume model using the rest frame temperature and chemical potential. The fits are to $\pi^{-}$, $\pi^{+}$, and combined $\pi^{+}+\pi^{-}$ spectra from the most central heavy ion collisions, as measured by the HADES~\cite{HADES1}, STAR~\cite{STAR1,STAR2,STAR3}, PHENIX~\cite{PHENIX1}, and ALICE~\cite{ALICE1,ALICE2,ALICE3,ALICE4} Collaborations.}
\label{fig7}
\end{figure*}

\begin{figure}[htbp]
\minipage{0.49\textwidth}
\hspace{-0.09cm}
\includegraphics[width=\textwidth]{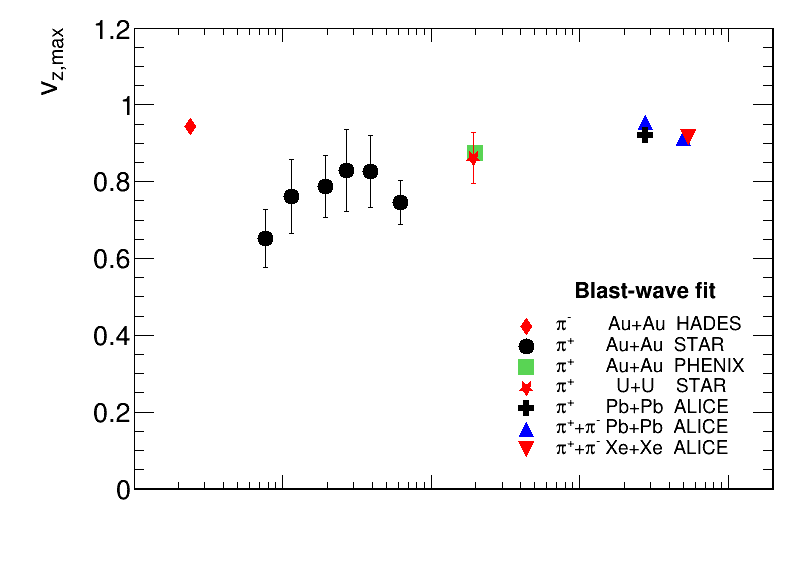}
\endminipage\hfill
\vspace*{-0.9cm}
\minipage{0.49\textwidth}
\includegraphics[width=\textwidth]{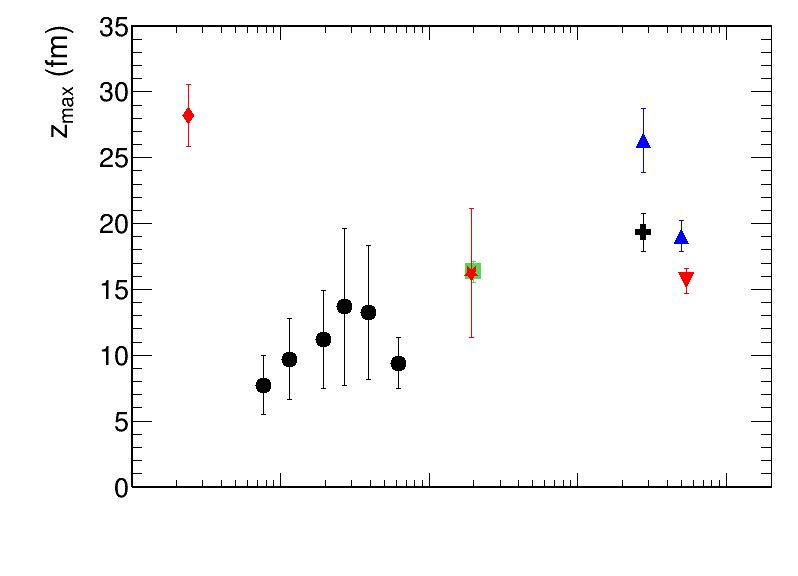}
\endminipage\hfill
\vspace*{-0.92cm}
\minipage{0.49\textwidth}
\hspace{-0.33cm}
\includegraphics[width=\textwidth]{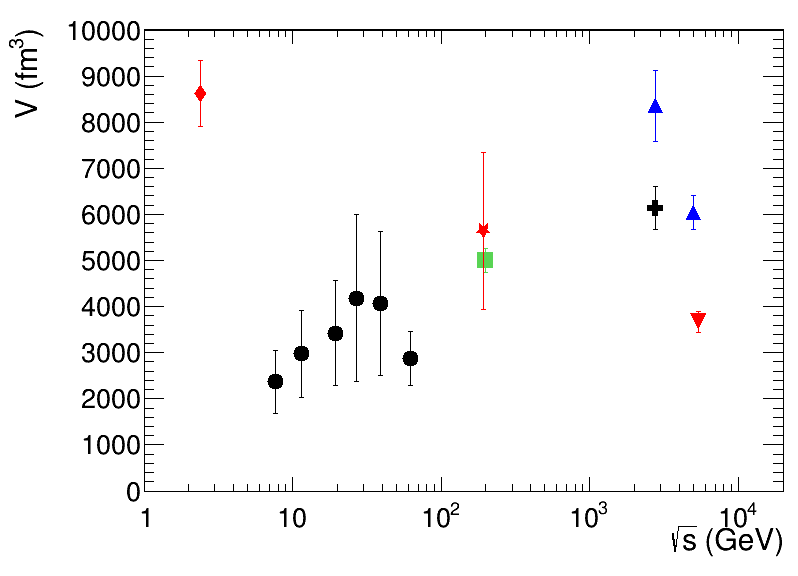}
\endminipage\hfill
\hspace{0.0\textwidth}
\vspace{-0.3cm}
\caption{(Color online) Energy dependence of maximum longitudinal velocity $v_{z,\mathrm{max}}$, maximum fire cylinder half length $z_{\mathrm{max}}$, and total volume $V$, derived from fit parameters using the cylindrically symmetric finite volume Boltzmann–Gibbs blast-wave model incorporating Planck transformed temperature and chemical potential.}
\label{fig8}
\end{figure}

Figure~\ref{fig3} presents the transverse momentum spectra of positively charged $\pi^+$ mesons, obtained by the STAR Collaboration~\cite{STAR1} in Au+Au collisions at $\sqrt{s_{NN}}= 7.7, \ 11.5, \ 19.6, \ 27 \ \text{and} \ 39$ GeV, within the rapidity range $-0.1 < y < 0.1$ and the 0--5$\%$ centrality class. The solid and dashed curves represent fits to the experimental data using Maxwell–Boltzmann transverse momentum distributions from the cylindrically symmetric BGBW model. The solid curve corresponds to the finite volume fire cylinder $(\eta_{\max} < \infty)$ with Planck transformed temperature and chemical potential (Eq.~\eqref{t12a}), while the dashed curve corresponds to the infinite volume fire cylinder $(\eta_{\max} = \infty)$ with rest frame temperature and chemical potential (Eq.~\eqref{t9a}).
The best fit parameters for the finite volume BGBW model [Eq.~(\ref{t12a})] and the infinite volume BGBW model [Eq.~(\ref{t9a})] are listed in Tables~\ref{t1} and \ref{t2}, respectively. Experimental data points are denoted by symbols, with numbers indicating corresponding scaling factors applied for visual clarity of the data. For Au+Au collisions, the freeze-out fire cylinder radius is $R_c = 6.98$~fm. The curves of these functions practically coincide and provide a good description of the experimental data in the experimentally accessible momentum range. However, the extracted parameters differ significantly (see Tables~\ref{t1} and \ref{t2}) which indicates the difference at the asymptotics of two functions mainly at the region for very low momenta particle production. The infinite volume BGBW model demonstrates higher rest frame temperature and chemical potential than the Planck transformed values from the finite volume BGBW model. This agrees with the general theory, which predicts that the Planck transformed temperature and chemical potential are reduced relative to their values in the rest frame (see, e.g., Refs.~\cite{Haar,Parvan2024}).

Figure~\ref{fig4} displays the transverse momentum spectra of $\pi^+$ mesons obtained by the STAR Collaboration in Au+Au collisions at $\sqrt{s_{NN}} = 62.4$ GeV~\cite{STAR2} and U+U collisions at $\sqrt{s_{NN}} = 193$ GeV~\cite{STAR3}, within the rapidity range $-0.1 < y < 0.1$ and the 0--5$\%$ centrality class. The figure also includes PHENIX data on $\pi^+$ mesons for Au+Au collisions at $\sqrt{s_{NN}} = 200$ GeV~\cite{PHENIX1} in the 0--5$\%$ centrality class at mid-rapidity. The solid and dashed curves represent fits to the experimental data using Maxwell–Boltzmann transverse momentum distributions from the cylindrically symmetric BGBW model. The solid curve corresponds to the finite volume fire cylinder $(\eta_{\max} < \infty)$ with Planck transformed temperature and chemical potential (Eq.~\eqref{t12a}), while the dashed curve corresponds to the infinite volume fire-cylinder $(\eta_{\max} = \infty)$ with rest frame temperature and chemical potential (Eq.~\eqref{t9a}). Both distributions are integrated over the rapidity interval $-0.1 < y < 0.1$ for all experimental data, including the PHENIX measurements. The freeze-out fire-cylinder radius is $R_c = 6.98$~fm for Au+Au collisions and $R_c = 7.44$~fm for U+U collisions. Experimental data points are denoted by symbols, with numbers indicating applied scaling factors. The curves of these functions practically coincide and provide a good description of the experimental data. The best fit parameters for the finite volume BGBW model and the infinite volume BGBW model are summarized in Tables~\ref{t1} and \ref{t2}, respectively. The infinite volume BGBW model yields a higher rest frame chemical potential than the Planck transformed chemical potential from the finite volume BGBW model. Conversely, its rest frame temperature is lower than the corresponding Planck transformed temperature from the finite volume model. This temperature behavior needs further theoretical explanation.

Figure~\ref{fig5} shows the transverse momentum spectra of charged $\pi^+ + \pi^-$ and $\pi^+$ mesons obtained by the ALICE Collaboration~\cite{ALICE1,ALICE2} in Pb+Pb collisions at $\sqrt{s_{NN}} = 2.76$ TeV, within the the 0--5$\%$ centrality class, and rapidity ranges $-0.8 < y < 0.8$ and $-0.5 < y < 0.5$, respectively. The solid and dashed curves represent fits to the experimental data using Maxwell–Boltzmann transverse momentum distributions derived from the cylindrically symmetric BGBW model. The solid curve corresponds to the finite volume BGBW model with Planck transformed temperature and chemical potential [Eq.~\eqref{t12a}], whereas the dashed curve corresponds to the infinite volume BGBW model with rest frame temperature and chemical potential [Eq.~\eqref{t9a}]. Both distributions are integrated over the rapidity intervals $-0.8 < y < 0.8$ for $\pi^+ + \pi^-$ and $-0.5 < y < 0.5$ for $\pi^+$. Experimental data are denoted by symbols, with numbers indicating the scaling factors applied for visual clarity of the data. For Pb+Pb collisions, the freeze-out fire cylinder radius is $R_c = 7.11$~fm. The two curves practically coincide and provide a good description of the experimental $\pi^+$ distribution. However, the transverse momentum distributions from the BGBW model fail to describe the combined $\pi^+ + \pi^-$ distribution. Consequently, for this $p_T$ distribution, the fit was restricted to the range $p_T < 3.1$~GeV. The best fit parameters for the finite volume BGBW model [Eq.~\eqref{t12a}] and the infinite volume BGBW model [Eq.~\eqref{t9a}] are summarized in Tables~\ref{t1} and \ref{t2}, respectively. The infinite volume BGBW model demonstrates higher rest frame temperature and chemical potential than the corresponding Planck transformed values obtained with the finite volume BGBW model. This agrees with the general theory of relativistic thermodynamics~\cite{Haar,Parvan2024}.

Figure~\ref{fig6} displays the transverse momentum spectra of charged pions $\pi^+ + \pi^-$ measured by the ALICE Collaboration in Pb+Pb collisions at $\sqrt{s_{NN}} = 5.02$ TeV~\cite{ALICE3} and Xe+Xe collisions at $\sqrt{s_{NN}} = 5.44$ TeV~\cite{ALICE4}, within the rapidity range $-0.5 < y < 0.5$ and the 0--5$\%$ centrality class. The figure also includes $\pi^-$ spectra in Au+Au collisions at $\sqrt{s_{NN}} = 2.4$ GeV measured by the HADES Collaboration in the 0--10$\%$ centrality class at mid-rapidity~\cite{HADES1}. The solid and dashed curves represent fits to the experimental data using Maxwell–Boltzmann transverse momentum distributions derived from the cylindrically symmetric BGBW model. The solid curve corresponds to the finite volume fire-cylinder ($\eta_{\max} < \infty$) with Planck transformed temperature and chemical potential [Eq.~\eqref{t12a}], whereas the dashed curve corresponds to the infinite volume fire-cylinder ($\eta_{\max} = \infty$) with rest frame temperature and chemical potential [Eq.~\eqref{t9a}].
Experimental data points are denoted by symbols, with numbers indicating the applied scaling factors applied for visual clarity of the data. The freeze-out fire-cylinder radius is $R_c = 6.98$~fm for Au+Au collisions, $R_c = 7.11$~fm for Pb+Pb collisions, and $R_c = 6.11$~fm for Xe+Xe collisions.  The two curves practically coincide. The transverse momentum distributions from the BGBW model fail to describe the combined $\pi^+ + \pi^-$ distributions obtained by the ALICE Collaboration. Consequently, for these $p_T$ distributions, the fit was restricted to $p_T < 3.1$~GeV. To achieve the best fit for the HADES data, the range was restricted to $p_T < 0.57$~GeV. The best fit parameters for the finite volume BGBW model [Eq.~\eqref{t12a}] and the infinite volume BGBW model [Eq.~\eqref{t9a}] are summarized in Tables~\ref{t1} and \ref{t2}, respectively. For the ALICE data, the infinite volume BGBW model yields higher rest frame temperature and chemical potential than the corresponding Planck transformed values from the finite volume model. This agrees with the general theory of relativistic thermodynamics~\cite{Haar,Parvan2024}. In contrast, the HADES data at low collision energy are not well described by the infinite volume BGBW model with rest frame temperature and chemical potential [Eq.~\eqref{t9a}].

\begin{figure}[htbp]
\includegraphics[width=0.47\textwidth]{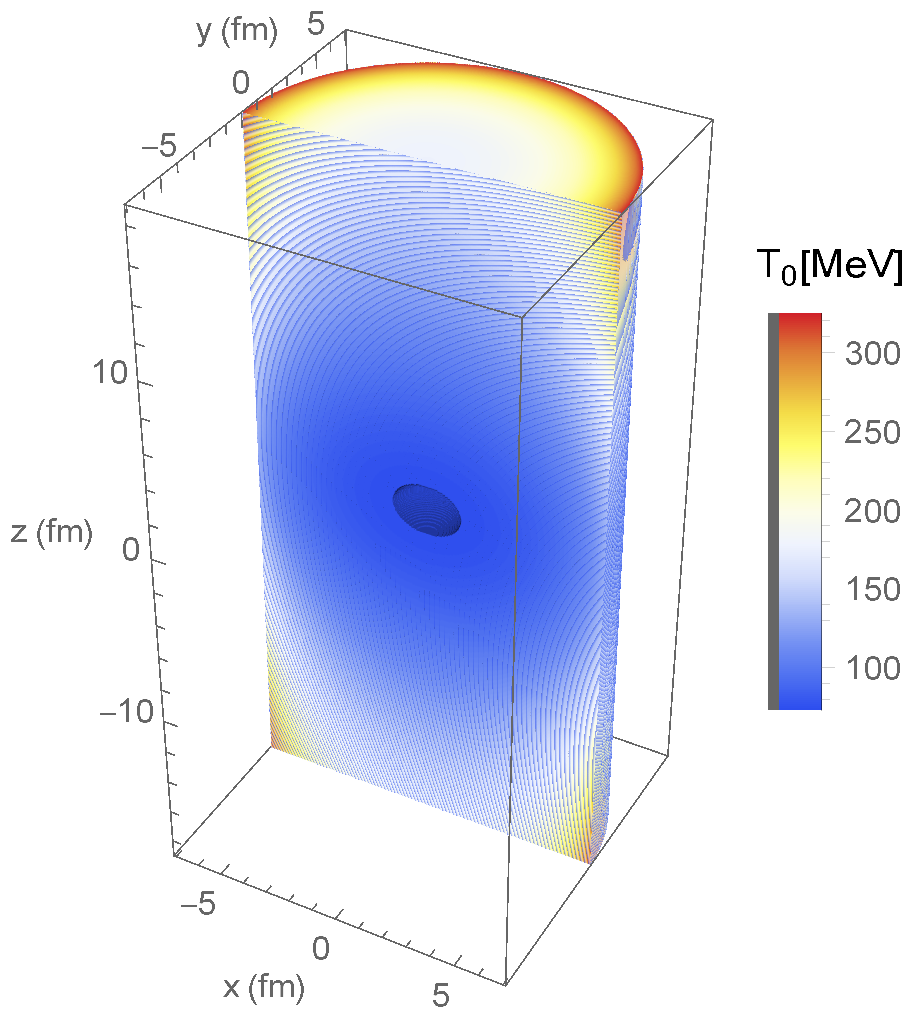}
\caption{(Color online) Distribution of the rest frame temperature $T_0$ of fluid elements at freeze-out for the fire cylinder model. Calculations correspond to combined $\pi^+ + \pi^-$ transverse momentum distributions measured by the ALICE Collaboration~\cite{ALICE3} in central Pb+Pb collisions at $|y| < 0.5$ and center-of-mass energy $\sqrt{s_{NN}} = 5.02$ TeV.}
\label{fig9}
\end{figure}

The cylindrically symmetric BGBW model with Planck transformation accurately reproduces the experimental data on particle production at STAR and PHENIX energies. However, it fails to describe the hard part of the spectra of high energy ALICE data, necessitating a fitting range restriction to $p_T < 3.1$ GeV. Similarly, for HADES data, the fitting range was constrained to maintain accuracy due to uncertainties at large $p_T$.

Figure~\ref{fig7} shows the energy dependence of the fit parameters obtained using the cylindrically symmetric BGBW model. The fits are for $\pi^{-}$, $\pi^{+}$, and $\pi^{+}+\pi^{-}$ mesons produced in the most central $A$--$A$ collisions over the energy range $2.4$~GeV $\leqslant \sqrt{s_{NN}} \leqslant 5.44$~TeV. Solid and open symbols represent fits to experimental data using Maxwell-Boltzmann transverse momentum distributions from the model. Solid symbols correspond to the finite volume fire-cylinder ($\eta_{\max} < \infty$) with Planck transformed temperature and chemical potential (Eq.~\eqref{t12a}), while open symbols corresponds for the infinite volume fire cylinder ($\eta_{\max} = \infty$) with the rest frame temperature and chemical potential (Eq.~\eqref{t9a}). In the finite volume BGBW model, the laboratory frame temperature remains lower than the rest frame temperature of fluid elements in the infinite volume BGBW model across the full heavy ion collision energy range, except for Au+Au collisions at $\sqrt{s_{NN}} = 200$~GeV~\cite{PHENIX1} and U+U collisions at $\sqrt{s_{NN}} = 193$~GeV~\cite{STAR3}. These two temperatures follow notably different trends. In the finite volume model, the laboratory frame temperature rises gradually with center-of-mass energy up to 200 GeV before declining slightly across the LHC energy regime. In contrast, the rest frame temperature in the infinite volume model decreases with energy up to 200 GeV and then rises slightly at the LHC energies. The laboratory frame chemical potential in the finite volume BGBW model is consistently lower than the rest frame chemical potential in the infinite volume BGBW model across the entire heavy ion collision energy range. They differ by a fixed value. Nevertheless, both chemical potentials exhibit identical dependence on collision energy. The chemical potentials increase with energy up to top RHIC energies and then decrease slightly at LHC energies. The transverse flow velocities in both models are significant and follow nearly identical trends, increasing monotonically with the collision energy. Nevertheless, they differ numerically between the two models. The maximum spacetime rapidity in the infinite volume BGBW model is infinite, $\eta_{\mathrm{max}} = \infty$. In the finite volume BGBW model, the maximum spacetime rapidity is finite and increases with collision energy, except for Au+Au collisions at HADES energy. In the laboratory frame, both the temperature and chemical potential are reduced relative to their rest-frame values, as required by the Planck transformations. Surprisingly, however, at $\sqrt{s_{NN}} = 193$ and $200$ GeV, the extracted rest-frame temperature falls below the laboratory frame temperature. This apparent reversal constitutes an anomaly that warrants further investigation. It should be emphasized that this temperature behavior in the finite and infinite volume BGBW models explains why the extracted temperature in the conventional blast-wave model is typically higher than that obtained from standard statistical models based on Tsallis non-extensive statistics~\cite{Cleymans13,Parvan2017,Azmi20}. This difference arises because, in the usual Tsallis distribution, the temperature is defined in the laboratory frame, whereas in the conventional blast-wave model it is defined in the local rest frame of the fluid element. The laboratory frame temperature in the finite volume BGBW model is in good agreement with the temperature obtained from the standard Tsallis distribution.

Figure~\ref{fig8} shows the energy dependence of the maximum longitudinal flow velocity $v_{z,\max}$, the maximum fire cylinder half length $z_{\max}$, and the freeze-out volume $V$, as obtained from the fit parameters of the cylindrically symmetric finite volume BGBW model incorporating Planck transformed temperature and chemical potential. The maximum longitudinal velocity is calculated as $v_{z,\max} = \tanh\eta_{\max}$ using Eq.~(\ref{t14}). The freeze-out volume $V$ and the fire-cylinder half length $z_{\max}$ are determined from Eqs.~(\ref{t7}) and (\ref{t8}), respectively. Numerical results for these quantities are presented in Table~\ref{t3}. In the most central collisions, the maximum longitudinal velocity $v_{z,\max}$ increases with center-of-mass energy, approaching very close to the speed of light limit at the energies investigated by ALICE. This reflects a rapid longitudinal expansion of the fire cylinder with increasing collision energy. As a result, both the freeze-out volume $V$ and the fire-cylinder half length $z_{\max}$ become very large. Nevertheless, these values remain physically reasonable when compared to the combined volume of two colliding nuclei in their ground state. For example, the volume of the system in the ground state is $V_0 = 2851.9$~fm$^3$ for Au+Au, $3445.4$~fm$^3$ for U+U, $3011.1$~fm$^3$ for Pb+Pb, and $1910.9$~fm$^3$ for Xe+Xe collisions. Both the freeze-out volume $V$ and the fire cylinder half length $z_{\max}$ increase with center-of-mass energy, exhibiting similar trends. This behavior is expected because the transverse radius of the fire-cylinder is held fixed in our model, making $V$ and $z_{\max}$ scale proportional with the longitudinal extent.

Figure~\ref{fig9} displays the spatial distribution of the rest frame temperature $T_0$ of fluid elements at freeze-out for the cylindrically symmetric finite volume BGBW model incorporating Planck transformed temperature and chemical potential. Calculations correspond to combined $\pi^+ + \pi^-$ transverse momentum distributions obtained by the ALICE Collaboration~\cite{ALICE3} in central Pb+Pb collisions at $|y| < 0.5$ and center-of-mass energy $\sqrt{s_{NN}} = 5.02$ TeV. Applying the Planck boost transformation (Eq.~\eqref{6}) together with Eqs.~\eqref{t11a} and \eqref{t5}, the proper temperature and chemical potential at freeze-out in the local rest frame of the fluid element are given by
\begin{align}\label{z1}
  T_{0}(r,\eta_{\|},\phi) &=\frac{T \cosh\eta_{\|}}{\sqrt{1-\bar{v}_{T}^{2}(r)}}, \nonumber \\
  \mu_{0}(r,\eta_{\|},\phi) &=\frac{\mu \cosh\eta_{\|}}{\sqrt{1-\bar{v}_{T}^{2}(r)}},
\end{align}
where $  T  $ and $  \mu  $ denote the temperature and chemical potential in the laboratory frame. Position four-vector of a fluid element on the freeze-out hypersurface (the “fire-cylinder”) is parameterized as
\begin{equation}\label{z2}
  x^{\mu} = (\tau_{\mathrm{f}}\cosh\eta_{\|},r\cos\phi,r\sin\phi,\tau_{\mathrm{f}}\sinh\eta_{\|}),
\end{equation}
with the freeze-out proper time $\tau_\mathrm{f} = R_c / \bar{v}_{T\mathrm{f}}$, radial coordinate $0 \leqslant r \leqslant R_c$, azimuthal angle $0 \leqslant \phi < 2\pi$, and space-time rapidity bounded by $|\eta_{\|} | \leqslant \eta_{\max}$. The numerical values of the freeze-out parameters are $T=72.9$ MeV, $\bar{v}_{T\mathrm{f}}=0.836$, $\eta_{\max}=1.545$ and $R_c = 7.11$ fm. The maximum longitudinal velocity is $v_{z,\max}=0.913$. As shown in Fig.~\ref{fig9}, despite a uniform laboratory frame temperature $T$ across the fire cylinder, the proper temperature $T_0$ of the fluid elements is inhomogeneous and anisotropic. The proper temperature $T_0$ increases with the fluid element’s velocity in the laboratory frame, reaching its maximum on the outer edges of the fire cylinder, where velocities of elements are the highest.

\section{Summary and conclusion}
In this study, we presented results for the Boltzmann–Gibbs blast-wave models with spherical and cylindrical symmetry with a finite volume nuclear fireball, in which the laboratory frame temperature and chemical potential are derived from their rest frame values using Planck transformation. Our results reveal that the geometric symmetry of the freeze-out hypersurface (spherical or cylindrical) notably alters the shape of the transverse momentum ($p_T$) distributions. We demonstrate that the transverse momentum spectra predicted by the finite volume blast-wave model with cylindrical symmetry systematically differs from those of the conventional infinite volume blast-wave model, with the discrepancy increasing with particle rapidity. This highlights the critical importance of taking into account the finite fireball volume, particularly at large particle rapidities, where the boost invariance may not be conserved any more. Incorporating a finite volume into the cylindrically symmetric blast-wave model enables us to extract key physical parameters of the fireball -- namely, its total spatial volume, maximum longitudinal flow velocity, maximum half-length of the fire cylinder, and maximum spacetime rapidity which remain inaccessible in the conventional infinite volume blast-wave framework. The finite volume blast-wave model with cylindrical symmetry produces Lorentz-invariant $p_T$ spectra that depends on the particle’s rapidity $y$. This contrasts with the conventional infinite volume blast-wave model, which, by construction, is boost invariant and therefore exhibits no rapidity dependence in the transverse momentum distribution.

The cylindrically symmetric finite volume BGBW model offers several important advantages over the conventional blast-wave model: \\
1) It allows the fireball volume to be calculated consistently within the model, without any additional assumptions. This is impossible in the conventional blast-wave model, where the fireball volume is infinite by definition; \\
2) The temperature and chemical potential can be specified directly in the laboratory frame (in which the fireball is at rest). The local temperature and chemical potential in the rest frame of each fluid element are then obtained as spatially varying fields using the Planck transformations; \\
3) Conversely, the temperature and chemical potential can be specified in the local rest frame of each fluid element, after which the corresponding quantities in the laboratory frame are obtained as spatially dependent fields via the Planck transformations.

We compared the cylindrically symmetric finite volume Boltzmann–Gibbs blast-wave model -- using Planck transformed laboratory frame parameters $  T  $ and $  \mu  $ -- with the conventional infinite volume blast-wave model, which employs rest frame parameters $  T_0  $ and $  \mu_0  $. Both models were fitted to the transverse momentum spectra of charged pions produced in the most central heavy-ion collisions over a broad center-of-mass energy range, $  \sqrt{s_{NN}} = 2.4  $~GeV to $5.44$~TeV. In both cases, the energy dependence of the temperature, chemical potential, and transverse flow velocity was studied at a fixed radius of the fire cylinder.
We find that the rest frame temperature and chemical potential extracted from the conventional infinite volume blast-wave model are systematically higher than the corresponding Planck transformed laboratory frame values obtained from the finite volume BGBW model. This result is consistent with expectations from relativistic thermodynamics. An exception is observed at $  \sqrt{s_{NN}} = 193  $ and $200$~GeV, where the rest frame temperature falls slightly below the laboratory frame value.
These findings explain why the temperature extracted in the conventional blast-wave model is typically higher than that obtained from statistical models based on Tsallis non-extensive statistics. The systematic difference arises because temperature in the Tsallis distribution is defined in the laboratory frame, whereas in the conventional blast-wave model it is defined in the local rest frame of each fluid element.

Furthermore, using the cylindrically symmetric finite volume Boltzmann–Gibbs blast-wave model together with the Planck transformations, we have calculated the energy dependence of the fire cylinder volume as well as other key quantities at kinetic freeze-out, including the maximum longitudinal flow velocity, the maximum half-length of the fire cylinder, and the maximum space-time rapidity. These quantities remain finite and reflect the realistic dimensions of the systems created in heavy-ion collisions.

Conversely, in the conventional infinite volume blast-wave model, the fire cylinder volume, the maximum half-length of the fire cylinder, and the maximum space-time rapidity are infinite at all collision energies, while the maximum longitudinal flow velocity equals the speed of light. These non-physical results are incompatible with the realistic description of the system created in heavy-ion collisions.

In both models, the system expands transversely according to Hubble’s law. In the finite volume BGBW model, however, the system undergoes rapid expansion in both the longitudinal and transverse directions, with all expansion velocities remaining finite and well below the speed of light.

We also find that the classical blast-wave model fails to describe the hard part of the transverse momentum spectra of produced particles at both very low (HADES) and very high (ALICE) collision energies. To improve the description of the experimental data, the present framework can be extended by replacing the classical Boltzmann–Gibbs statistical particle production on the blast-wave surface at freeze-out with non extensive Tsallis statistical production under the same conditions.

\vskip0.1in
\noindent
{\bf Acknowledgments}

This work was supported in part by the RSCF grant, N22-72-10028-\foreignlanguage{russian}{П} and the Romanian Ministry of Research, Innovation and Digitalization, Project PN 23 21 01 01/2023.

\end{document}